\documentclass[hyper,a4paper,11pt,oneside]{JHEP3}

\JHEPspecialurl{http://jhep.sissa.it/JOURNAL/JHEP3.tar.gz}

\usepackage{graphicx}
\usepackage{dcolumn}
\usepackage{bm}
\usepackage{epsfig,multicol,bbm}

\newcommand\fverb{\setbox\fverbbox=\hbox\bgroup\verb}
\newcommand\fverbdo{\egroup\medskip\noindent%
            \fbox{\unhbox\fverbbox}\ }
\newcommand\fverbit{\egroup\item[\fbox{\unhbox\fverbbox}]}
\newbox\fverbbox

\title{Contributions of Riemann invariants to the Entropy of Extremal
Black Holes.}

\author{A. Ulacia Rey \\ {Instituto de Cibern\'etica Matem\'atica y F\'{\i}sica
(ICIMAF). Calle E No-309 Vedado, cp-10400. La Habana, Cuba.} \\
E-mail: \email{alain@icmf.inf.cu;}}

\received{10/November/2008}       
\accepted{\today}       

\preprint{\hepth{0811.2371}}  

\abstract{We use the entropy function formalism introduced by A. Sen
to obtain the entropy of $AdS_{2}\times S^{d-2}$ extremal and static
black holes in four and five dimensions, with higher derivative
terms of a general type. Starting from a generalized
Einstein--Maxwell action with nonzero cosmological constant, we
examine all possible scalar invariants that can be formed from the
complete set of Riemann invariants (up to order 10 in derivatives).
The resulting entropies show the deviation from the well known
Bekenstein--Hawking area law $S=A/4G$ for Einstein's gravity up to
second order derivatives.}


\begin{document}
\section{Introduction}
 The idea of the ``attractor mechanism'' was firstly applied in
the context of supergravity. Later it was reinterpreted to calculate
the entropy of extremal black holes (BH)\cite{Sen:2007qy}. Many
articles already have been published in which this method has been
used successfully in different contexts
\cite{Examples_EF_1},\cite{Examples_EF_2},\cite{Examples_EF_3}. The
popularity of this idea possibly follows from the simplicity of the
entropy function method. The fact that the geometry near the horizon
implies that all scalar fields and sources take constant values, is
perhaps, among its properties, what makes this method easier to
apply. This idea is consistent with the so--called ``attractor
mechanics'' for supersymmetric backgrounds, in which BH
configurations near the horizon depend only on the electric and
magnetic charge carried by the BH, and not on the asymptotic values
of the corresponding scalar field \cite{Sen:2007qy}. Indeed, by
using the entropy function mechanism, the entropy in the proximity
of the BH horizon can be regarded as the Legendre transformation of
specific suitable parameters. Since the entropy will depend on these
charges, it is important to emphasize that the entropy function
formalism computes the entropy of an extremal BH as the entropy of a
non--extremal BH in the limit as we approach the horizon.

In $d$ dimensions and near the horizon, the geometry of a
spherically symmetric and extremal BH implies that the isometry
group has the $SO(1,2)\times SO(d-1)$ form
\cite{James1},\cite{James2},\cite{Astefanesei1}. This follows from
the fact that near the horizon the metric has the
Robinson--Bertotti(RB) form $AdS_{2}\times S^{d-2}$
\cite{Robinson},\cite{Bertotti}. If we assume that the addition of
higher derivative terms in the action do not destroy the RB geometry
near the horizon, then we could consider this metric as valid in
general, even for dimensions larger than four. It is worth noting
that the RB solution is not asymptotically flat, as it is the
product of two spaces~\cite{T.Ortin}: the anti-de Sitter $AdS$ space
and the sphere $S$. In $d=4$ both are two--dimensional spaces with
the same curvature constants, $2/R^{2}_{AdS}$, but with different
sign, with the $AdS$ space having negative curvature and being
invariant with respect to the isometry group $SO(1,2)$, while
$S^{2}$ is the two--dimensional sphere with positive curvature and
invariant with respect to the group $SO(3)$. Since the RB group of
symmetry is bigger than the isometry group $SO(1,1)\times SO(3)$ of
the extremal Reissner Nordstrom (RN), the consequent symmetry
increase makes it possible to consider the RB space as a vacuum
state for this theory, similar to a Minkowski vacuum \cite{Gibbons}.
In this sense, the extremal RN black hole at infinity is
asymptotically flat ({\it i.e.} Minkowski), but near the horizon the
geometry is described by the RB metric. Therefore, the extremal RN
BH can be seen as a gravitational soliton. This interpretation is
closely related to the recent discovery of AdS/CFT correspondence,
in which the $AdS$ gravity can be linked with gauge theories defined
on the $AdS$ boundary.

In this paper we obtain the entropy of extremal and static BH taking
into account higher derivative terms. The inclusion of these terms
in theories of high order gravity can be done, for example, using
the terms of Gauss-Bonnet(GB)
\cite{Morales:2006gm},\cite{Astefanesei2},\cite{Astefanesei3}. These
terms could also appear in various situations, such as string
theories,
branes\cite{Maldacena:1996ky},\cite{Duff:1999rk},\cite{Peet:2000hn}
and semiclassical quantum gravity. In particular, higher derivative
terms appear in string theories when the effective low-energy limit
is evaluated. It is well-known that in higher order gravity problems
associated with renormalization may occur. In the particular case of
GB gravity, the action formed by three second order invariants
insure the elimination of these renormalization problems, though
other problems come up, such as the emergence of ghost fields
($i.e.$ massive particle with spin two). The discussion of these
problems is beyond the scope of this article, though a good review
on these issues can be found in \cite{Farhoudi} and
\cite{B.Zwiebach}.

 The main objective of this paper is to present a set of
approximate solutions for the entropy of extremal BH's near the
horizon of geometry in $d=4$ and $d=5$ dimensions. To this end we
used Sen's mechanism, the initial condition derived from the
Einstein--Maxwell theory with cosmological constant, as well as a
complete set of Riemann invariants. It is important to point out
that in the cases when the equations of motion near the BH horizon
cannot be solved analytically, we can attempt to construct an
iterative power series in the $w$ parameter inversely to the BH
charge.
 On the other hand, in order to include the Riemann set of invariants as
corrections of higher derivatives, we have to considere the complete
set of invariants defined by Carminati and McLenaghan (CM)
\cite{JCarminati}, plus the $m_{6}$ invariant introduced by Zakhary
and McIntosh \cite{Zakhary_McIntosh}.

The CM invariants are scalars built from the Riemann
tensor $R_{abcd}$, the Weyl tensor $C_{abcd}$ (and its dual) and
the trace-free Ricci tensor, defined as
$S_{ab}=R_{ab}-(1/d) R\,g_{ab}$ (in $d$-dimensions), leading to
six real scalars $R,r_{1},r_{2},r_{3},m_{3},m_{4}$ and five
complex scalars $w_{1},w_{2},m_{1},m_{2},m_{5}$, making a total of sixteen
invariant scalars (we notice that $r,w$ and $m$ are respectively associated with Riemann,
Weyl and mixed invariants). The CM set of scalars yields the
required number of invariants for the Einstein-Maxwell and perfect fluid
cases. Furthermore, by including the $m_{6}$ invariant, it has
been proven that the CM set becomes a {\it{complete set}}, as it
covers the 90 possible cases (6 Petrov types $\times$ 15 Segre
types)\cite{Zakhary_McIntosh}. In other words,
a complete set of invariants should contain (at least for $d=4$ spacetimes)
the already well-known physical invariants, as well as the geometric ones.

In Table \ref{Definicion_de_Invariantes} (see Appendix) we show the
definitions and results for the set of non--null invariants for the
case of $AdS_{2}\times S^{d-2}$ geometry in $d=4$ and $d=5$. It is
important to point out that for $d=5$ it is not possible to
calculate the complex Riemann invariants, as they are yet unknown.
Additional invariants may be required for a more general spacetime
like the $d=5$ case. The question of how many invariants are
necessary to obtain a {\it complete set}, for $d>4$ is an open
problem. Likewise, the invariants in Table
\ref{Tabla_de_Invariantes} are organized by their respective degree.
Following the definitions given in \cite{Zakhary_McIntosh}, the
$j-th$ invariant $I^{p}_{j}$ is called the invariant of $order$ $p$,
and then if other invariant exists that can be written in the form
$I^{p}_{j}I^{q}_{k}I^{r}_{l}$, it is said to be an invariant of
$order \,\,p-q-r$ and the sum $p+q+r$ is his $degree$. For example,
in Table \ref{Definicion_de_Invariantes} the invariant
$\mathfrak{Re}(m_{1})$, which denotes the real part of the invariant
$m_{1}$, is of $order\,\,1-2$ and has third-{\it degree}. Thus, the
invariant $I^{p}_{j}$ will be {\it independent} if it can not be
written in terms of other invariants, either of equal or lower
degree. Also, two invariants are said to be {\it equivalent} if they
can be written in terms of each other, or as the product of other
invariants of lower degree. The {\it equivalence} relations
(syzygies) are also given in Table-\ref{Tabla_de_Invariantes}. It is
easy to see that in $d=4$ and $d=5$ all the invariants can be
written in terms of the $R,\,r_{1}$ and $r_{2}$ invariants.

It is important to remark that all the non-zero invariants  are
taken into account in our calculations. The remaining null
invariants are not included in the Table
\ref{Definicion_de_Invariantes}-\ref{Tabla_de_Invariantes}.
 Naturally, an important issue may arise in these tables, namely: why there are no
invariants or higher order terms of a gauge theory given the fact that we have
started from an Einstein-Maxwell theory where a gauge is contained?
The answer to this question is that, for simplicity, we have not considered
 in this paper terms such as $(F_{\mu \nu}F^{\mu
\nu})^2,\,F^{\mu}_{\,\,\,\nu}\,F^{\nu}_{\,\,\rho}F^{\rho}_{\,\,
\sigma}F^{\sigma}_{\,\,\mu},\,R\,F^2$,..., nor covariant $F_{\mu
\nu}\Box F^{\mu \nu}$,..., nor invariants of forms. Hence, just the
purely gravitational sector of the theory is considered in the CM
set. All calculations were carried out with the tensor package
GRTensor running on the algebraic computing program Maple.

This paper is organized in six sections. In section II the
generalized theory for $d=4$ and $d=5$ is written with higher
derivative terms built from the set of Riemann invariants. The
analytic solution in $d=4$ for invariants of second degree is
obtained in section III. Approximate solutions for $d=4$ of the
complete set of Riemann invariants are also examined in section III.
Analytic and approximate solutions are provided in section IV for
$d=5$. The generic GB case is shortly treated in section V. We
present our conclusion in section VI, while the Appendix provides
the set of Riemann invariants for an extreme BH background.

\section{Generalized theory in 4 and 5 dimensions.} We consider in
this section a higher order theory of gravity by introducing the
complete set of non--null Riemann invariants as the higher
derivative terms of the theory. We construct then the
Einstein--Maxwell action with cosmological constant and the
additional higher derivative terms like,

\begin{equation}      {\mathcal{S}}=\frac{1}{16\pi
      G_{d}}\int{dx^{d}\sqrt{-g}\left(R+\Lambda-\frac{F^{2}}{4}+{\mathcal{L}}^{d}_{inv}\right)},  \label{AccionGeneralizada}\end{equation}
where:
   \begin{eqnarray}
      {\mathcal{L}}^{d=4}_{inv}&=&a_{2}R^2+b_{2}R_{2}+a_{3}R^{3}+b_{3}R\,R_{2}+a_{4}R^4+b_{4}{R^{2}_{2}}+c_{4}{R^2}\,R_{2}+a_{5}R^{5}+
      b_{5}{R^{3}}\,R_{2}\,, \nonumber \\
      {\mathcal{L}}^{d=5}_{inv}&=&a_{2}R^2+b_{2}R_{2}+a_{3}R^{3}+b_{3}{R}\,R_{2}+c_{3}R_{3}+a_{4}R^4+b_{4}{R^{2}_{2}}+c_{4}{R^2}\,{R_{2}}+
      e_{4}{R}\,{R_{3}}\,.\nonumber \label{AccionGeneralizada5D}
   \end{eqnarray}
with $G_{d}$ being the $d$--dimensional Newton constant, $R$ the
Ricci scalar, $\Lambda$ the cosmological constant, $F_{\mu \nu}$ the
electromagnetic tensor, and $F^{2}=F_{\mu \nu}F^{\mu \nu}$, $R_{2}$
and $R_{3}$ the two first real Riemann invariants defined in Table
\ref{Definicion_de_Invariantes} for the metric
(\ref{Metrica_Extremal_en_4Dand5D}). The parameters
$a_{i},b_{i},c_{i},e_{i}...$ are the coupling constants for each
higher derivative term of i--th degree. The label
${\mathcal{L}}^{d}_{inv}$ denotes the higher derivative terms with
which we will shall work. In both cases $d=4$ and $d=5$ the
invariant terms inside ${\mathcal{L}}^{d}_{inv}$ form a complete
set, therefore we added all possible higher derivative terms until
the highest degree \footnote{ By ``the highest degree'' we refer to
the highest degree within the complete set of Carminati-McLenagham
invariants inside the sector of pure gravity.}. The most general
spacetime for a static and extremal BH with $AdS_{2}\times S^{d-2}$
topology near the horizon of geometry is:
    \begin{eqnarray}
      ds^{2}&=&v_{1}(-r^2dt^{2}+\frac{dr^{2}}{r^2})+v_{2}d{\Omega_{d-2}}, \label{Metrica_Extremal_en_4Dand5D}\\
        e^{a_{I}\Psi_{I}}|_{H}&=&u_{I}, \, \, \, {F}^{d}_{0r}=e, \, \, \,
          {F}^{d=4}_{\theta\phi}=p \sin\theta,\,\,{F}^{d=5}_{\theta\phi}=0, \\
        d\Omega^{2}_{d-2}&=& d\theta^{2}_{1}+\sum^{d-2}_{i=2}\prod^{i-1}_{j=1}\sin^{2}\theta_{j} d\theta^{2}_{i}, \,\, 0\leq
      \theta_{i} \leq \pi, \, 0\leq \theta_{d-2} \leq 2\pi, \, (1\leq i\leq d-3).
    \end{eqnarray}
 where $e$ and $p$ are functions related to the electric and
 magnetic charges, while the $v_{1}$ and $v_{2}$ are functions connected
 with the BH throat. The constants $u_{i}$ are the values of the
 scalars fields $\Psi_{i}$ on the BH horizon, if these fields are present (we assume $u_{i}=0$).
 We shall follow in this paper the formalism of Sen~\cite{Sen:2007qy}, in which the entropy
 function $\mathcal{E}$ is defined as:
\begin{equation}
    \mathcal{E}(\vec{u},\vec{v},\vec{e},\vec{q},\vec{p})=2\pi(e_{i}q_{i}-f(\vec{u},\vec{v},\vec{e},\vec{p})).
\end{equation}
and $f(\vec{u},\vec{v},\vec{e},\vec{p})$ is the Lagrangian density
$\sqrt{{-det}\,{g}}{\mathcal{L}}$, evaluated near the horizon of
this geometry. All these parameters can be determined by extremizing
the entropy function:
\begin{equation}
 \frac{\partial{\mathcal{E}}}{\partial{u_{i}}}=\frac{\partial{\mathcal{E}}}{\partial{v_{j}}}=
   \frac{\partial{\mathcal{E}}}{\partial{e}}=0,  \qquad {i=1..N}, {j=1..2}. \label{R4_4}
\end{equation}
The last set of equations are the equations of motion near the
horizon of the extremal background
(\ref{Metrica_Extremal_en_4Dand5D}). Thus, the BH entropy at the
extremal limit follows after solving the system of equations
(\ref{R4_4}) and substituting these parameters in the entropy
function. The result shows that $S_{BH}/2\pi$ may be regarded as the
Legendre transform of the function $f(\overrightarrow{u},
\overrightarrow{v},\overrightarrow{e},\overrightarrow{p})$, with
respect to the variables $e_{i}$. In the following section the
second order contributions are obtained.
\section{Case d=4}
\subsection{Analytic Solutions for invariants of second degree in
$d=4$. }
 The calculation of second order contributions to the extremal BH entropy is
equivalent to considering invariants of second degree. On the basis of
the correspondence principle, the Gauss-Bonnet(GB) solutions should
be contained in these results \cite{Morales:2006gm}, as well as the
classic Reissner-Nordstrom (RN) solution that was obtained through
the area law.
 In this case, we should take $a_{i}=b_{i}=c_{i}=0$ for $i\geq 3$ in
(\ref{AccionGeneralizada}), and then the function
$f(\vec{v},\vec{e},\vec{p})$ will be,
  \begin{equation}
     f(v_{1},v_{2},e,p)={\int_{S^{2}}}{\sqrt{-g}\left(R+\Lambda-\frac{F^{2}}{4}+a_{2}R^2+b_{2}R_{2} \right)d\theta
     d\phi}.   \label{eRN}
  \end{equation}
while the entropy function will take the form:

\begin{eqnarray}
  \mathcal{E}(v_{1},v_{2},e,q,p)&=& \Biggl\{ 2\,q{e}G_{4}+
    2\left( 2-\,{\frac {{v_{1}}}{{v_{2}}}}-\,{\frac {v_{2}}{v_{1}}} \right){a_{2}}+ \left( -{\frac {v_{1}}{8v_{2}}}-{\frac{{v_{2}}}{{8v_{1}}}}
      -\frac{1}{4}\right) {b_{2}}+\nonumber \\
      &+& \left( {\frac {{{p}}^{2}}{2v_{2}}}-2-{v_{2}}\,\Lambda \right)\frac{v_{1}}{2}-{\frac {{v_{2}}\,{{e} }^{2}}
    {4v_{1}}}+{v_{2}} \Biggr\}\frac{\pi}{G_{4}}\,.
\end{eqnarray}

From the equations of motion (\ref{R4_4}) we obtain the system:
 \begin{eqnarray}
    2\,q-\,\frac{v_{2}\,e}{2v_{1}G_{4}}=0,\\
        \,\frac{v_{1}\Lambda}{2}-1+2\,{\frac {a_{2}}{v_{1}}}-2\,{\frac {{
         v_{1}}\,{a_{2}}}{{{v_{2}}}^{2}}}+{\frac {{b_{2}}}{8v_{1}}}-{\frac {{v_{1}}\,{b_{2}}}{8{v_{2}}^{2}}}+{\frac
            {{{e}}^{2}}{4v_{1}}}+{\frac {{v_{1}}\,{p}^{2}}{4{v_{2}} ^{2}}}=0,\\
            -{v_{2}}\,\frac{\Lambda}{2}-1+2\,{\frac {{v_{2}}\,{a_{2}}}{{{v_{1}}}^{2
           }}}-2\,{\frac {{a_{2}}}{{v_{2}}}}+{\frac {{v_{2}}\,{b_{2}}}{{ {8v_{1}}}^{2}}}-{\frac {{b_{2}}}{{8v_{2}}}}+{\frac
       {{v_{2}}\,{{e}}^{2}}{{{4v_{1}}}^{2}}}+{\frac
       {{p}^{2}}{4v_{2}}}=0.
  \end{eqnarray}

Notice that when we solve the system of equations above, all its
solutions can be written in terms of the function $v_{2}$ as
follows,

\begin{eqnarray}
 v_{1}&=&{\frac {{v_{2}}}{{v_{2}}\,\Lambda+1}}, \qquad
   q ={\frac{f}{{8G_{4}}}}, \qquad
   {e}={\frac{f}{2({v_{2}}\,\Lambda+1)}},\hspace{0.5cm} \label{Sol_2_Orden_d4} \\
 f&=&
 \sqrt{8\,(\Lambda{v_{2}}+2)v_{2}-2(16\,{a_{2}}+b_{2})\,({v_{2}}\Lambda+2)v_{2}\Lambda-4\,{{p}}^{2}}\,. \nonumber
\end{eqnarray}
and the entropy for an extremal and static BH (taking into account
invariants of second degree) will be:

\begin{equation}
   {S}_{BH}=\left( 1-{\frac { \left( 16\,{a_{2}}+{b_{2}} \right)
     \Lambda}{{4}}} \right)\,\frac{\pi v_{2}}{G_{4}}-{\frac {\pi
   \,{b_{2}}}{{2G_{4}}}}.
   \label{SBH_Second_Order_d4}
\end{equation}
The GB solution that was found by Morales and Samtleben in
\cite{Morales:2006gm} can then be obtained substituting
$b_{2}=-8\alpha$, and $a_{2}=\alpha/2$ in
(\ref{SBH_Second_Order_d4}), thus:
 \begin{equation}
   S_{GB}=(v_{2}+4\alpha)\frac{\pi}{G_{4}}. \label{GB_4D}
 \end{equation} where the parameter $\alpha$ is the GB coupling
constant. Similarly, if we take $a_{2}=b_{2}=0$ in
(\ref{SBH_Second_Order_d4}), the entropy becomes the well-known RN
black hole entropy in the extremal limit and with magnetic charge
$p$:
\begin{equation}
   S_{BH}|_{a_{2}=b_{2}=0}=\frac{\pi v_{2}}{G_{4}}\equiv S^{d=4}_{RN}.
   \label{RN_entropy}
\end{equation}
The variables $v_{1},e$ and $q$ become,
\begin{equation}
   v_{1}={\frac {{v_{2}}}{{v_{2}}\,\Lambda+1}},\qquad
    q={\frac{\sqrt{2\,(\Lambda{v_{2}}+2)v_{2}-\,{{p}}^{2}}}{{4G_{4}}}},\qquad
   {e}={\frac{\sqrt{2\,(\Lambda{v_{2}}+2)v_{2}-\,{{p}}^{2}}}{{v_{2}}\,\Lambda+1}}.\label{Sol_RN_Lambda}
\end{equation} where it is straightforward to see (from  (\ref{SBH_Second_Order_d4})) that the cosmological constant by itself
does not change the Bekestein-Hawking area law, since it needs to be
accompanied by higher derivative terms of at least of second degree. In
fact, the cosmological constant $\Lambda$ just changes the geometry of the throat
(see (\ref{Sol_2_Orden_d4}),(\ref{Sol_RN_Lambda})). However,
the constants $a_{2}$ and $b_{2}$ in (\ref{SBH_Second_Order_d4})
represent the deviation from this law. In the following section
approximated solutions for higher order gravity are obtained, as well as the $R^{3},R^{4}$
and $R^5$ contributions.

\subsection{Approximated Solutions for the complete set of invariants
in $d=4$.}\label{Section-4}
 Due to the non-linearity of Einstein--Maxwell equations it is very
difficult to find exact analytic solutions with higher derivative
terms. In most cases, some approximation methods must be employed or
solutions must be found numerically. If we consider the complete set
of invariants, we cannot find the solutions of the system of motion
equations (\ref{R4_4}) in explicit form. In order to solve this
problem we shall introduce the parameter $w$ and make suitable
expansions around it. Indeed, $w$ in ({\ref{AccionGeneralizada}})
can always be extracted from the coupling constants
$a_{i},b_{i},c_{i},e_{i}$ by rescaling. Though, we should take into
account that $w$ must be extracted with the appropriate order. Since
RN solutions (case $\mathcal{L}_{inv}=0$) are well known
analytically, the first coupling constants for invariants of higher
degree are $a_{2}$ and $b_{2}$. As a consequence: $a_{2},b_{2}
\Rightarrow w\,a_{2},w\,b_{2},...$ and in general, we have:
\begin{equation}
a_{l},b_{m},c_{n},e_{p}\Longrightarrow
w^{l-1}{a}_{l},w^{m-1}{b_{m}},w^{n-1}{c_{n}},w^{p-1}{e_{p}}, \qquad
l,m,n,p=2,3,4,5,...
\end{equation}
The expansion parameter $w$ can be considered as the inverse
function of the charge, which is zero to leading order. This is the
form in which $w$ should appear in the entropy function
$\mathcal{E}$. Since we are mainly interested in approximate
solutions, then we should build series expansions of the functions
$e,q$ and $v_{1}$. In general, any of these functions can be
expanded in series around the parameter $w$ as:

\begin{equation}
h(w)=\sum_{k=0}^{\infty}\frac{w^k}{k!} \left(\frac{\partial^{k}
h}{\partial w^k}\right)_{w=0}.\end{equation}

This expansion allows us to write:
\begin{eqnarray}
  &e& \simeq e_{0}+w\Delta e_{1}+w^2\Delta e_{2}+w^3\Delta
  e_{3}+...,\nonumber  \\
    &q& \simeq q_{0}+w\Delta q_{1}+w^2 \Delta q_{2}+w^3 \Delta
    q_{3}+..., \nonumber \\
    &v_{1}& \simeq v_{1\,0}+w\Delta v_{1\,1}+w^2\Delta v_{1\,2}+w^3 \Delta v_{1\,3}+...,   \label{Approach} \\
  &\mathcal{E}& \simeq \mathcal{E}_{0}+w \Delta \mathcal{E}_{1}+w^2 \Delta \mathcal{E}_{2}+w^3 \Delta
  \mathcal{E}_{3}+... \nonumber
\end{eqnarray}

where in (\ref{Approach}) we have taken $\Delta
h_{k}=(1/k!)({\partial^{k} h/\partial w^{k}})_{_ {\tiny w=0}}$.
Consequently, $w$ fixes the level of approximation, so that when
$w=0$ we have solutions of order zero that will be the RN set
(\ref{RN_entropy})-(\ref{Sol_RN_Lambda}). Finally, by including the
complete set of invariants (i.e. in (\ref{AccionGeneralizada}) all
terms are taken completely), or the contributions of higher
derivative terms up to fifth degree for extremal BH entropy, the
approximation at different levels will be:
\begin{equation}
  S_{BH}=
  S^{d=4}_{RN}+\left(S^{(2)}_{BH}+S^{(3)}_{BH}w+S^{(4)}_{BH}w^2+S^{(5)}_{BH}w^3
  \right){\it w}+{\it O(w^5)}, \hspace{1.5cm}
\end{equation}
where:
 \begin{eqnarray}
    S^{(2)}_{BH} &=& -\left({\left( 16\,{a_{2}}+{b_{2}}
    \right)\frac{\Lambda}{4}}{v_{2}}+{\frac {{b_{2}}}{{2}}}\right)\frac{\pi}{G_{4}}, \label{SHB_All_Order_d4(2nd_Degree)} \\
     S^{(3)}_{BH}&=& \left(12\Lambda^{2}{v_{2}}\,a_{3}+(\frac{3}{4}v_{2}\Lambda^{2}+2\Lambda+
     \frac{1}{v_{2}}){b_{3}} \right)\frac{\pi}{G_{4}},\label{Approach_3nd_4D}\\
       S^{(4)}_{BH}&=& -\Biggl\{{32\,{v_{2}}\,{\Lambda}^{3}}\,a_{4}+{\frac { \left(
       {v_{2}}\,\Lambda+2 \right) ^{3}\,{b_{4}}}{8{v_{2}}^{2}}}+ {\frac {2\Lambda\, \left( {v_{2}}\,\Lambda+2
       \right)  \left( {v_{2}}\,\Lambda+1 \right) \,{c_{4}}}{{v_{2}}}}+(16a_{2}+b_{2})\times \nonumber \\
       &\times& \left(\frac{\Lambda\,(v_{2}\Lambda+2)^2\,b_{3}}{16v_{2}}+v_{2}a_{3}\Lambda^3 \right)  \Biggl\} \frac{\pi}{G_{4}},\label{Approach_4nd_4D}\\
       {S}^{(5)}_{BH}&=& \Biggl\{80 v_{2}\Lambda^{2}{a_{5}}+(5v_{2}\Lambda^{2}+16\Lambda+
  \frac{12}{v_{2}}){b_{5}}+\,96v_{2}\Lambda^2a_{3}^2+4(\frac{6}{v_{2}}+3v_{2}\Lambda^{2}+8\Lambda)a_{3}b_{3}+\\ \nonumber
  &+&\frac{(3v_{2}\Lambda+4)(v_{2}\Lambda+2)^2}{8\Lambda v_{2}^2}b_{3}^2+(16a_{2}+b_{2})\left(4\Lambda^2a_{4}v_{2}+\frac{(v_{2}\Lambda+2)^2c_{4}}{4v_{2}}+
  \frac{(v_{2}\Lambda+2)^4b_{4}}{64\Lambda^2v_{2}^3}\right)\Biggr\}\frac{\pi\,\Lambda^2}{G_{4}}.
 \end{eqnarray}\label{SBH_All_Order_d4}
%
 If we consider an effective theory where just invariants of third degree
are required, then we should take in (\ref{AccionGeneralizada}),
(\ref{SBH_All_Order_d4}) the constants with values
$a_{j}=b_{j}=c_{j}=e_{j}=0$ for $j\geq 4$, and then the entropy
solutions for extremal BH at a linear approximation is simply:
$S_{BH}=S_{RN}+(S^{(2)}_{BH}+S^{(3)}_{BH}w){\it w+O({w^3})}$. The
same can be done for fourth degree. We can consider this approach
more than enough because the analytical solution of second degree
(\ref{SBH_Second_Order_d4}) is exactly reproduced in
(\ref{SHB_All_Order_d4(2nd_Degree)}). Note that each entropy
contribution of invariants of $i-th$ degree is labeled as $S^{(i)}$,
then the super-indices $(i)$ just indicate the degree of the
invariants that produced it and not the approximation order. Note
that we have not written the solutions for $e,q$ and $v_{1}$,
because they are too long to be written in a paper. In the set of
approximated solutions the non-linear terms with respect to the
coupling constants appear in $S^{(4)}_{BH}$ and $S^{(5)}_{BH}$ with
contributions like $a_{2}a_{3},a_{2}b_{3},b_{2}a_{3},b_{2}b_{3}$ and
$ a^{2}_{3},a_{3}b_{3},b^{2}_{3},a_{2}a_{4},a_{2}c_{4},
a_{2}b_{4},b_{2}a_{4},b_{2}c_{4},b_{2}b_{4}$ respectively.\\
\section{ Case d=5}
\subsection{Analytic Solutions for invariants of second degree in
$d=5$.} In this case, the function $f(\vec{v},\vec{e},\vec{p})$
takes the same form as (\ref{eRN}), but the integration is carried
on over $S^{3}$, hence the entropy function is given by:
\begin{eqnarray}
  \mathcal{E}(v_{1},v_{2},e,q,p)&=&\,2\,\pi \,qe-{\frac {{\pi }^{2}
    \left(3\,v_{1}-v_{2} \right) ^{2}a_{2}}{ G_{5}\,v_{1}\,\sqrt
     {v_{2}}}}-{\frac3{40}}\,{\frac {{\pi } ^{2} \left(
      2\,v_{1}+v_{2}\right) ^{2}b_{2}}{G_{5}\,v_{1}\,\sqrt
    {v_{2}}}}-,\,\nonumber \\
  &-&\,{\frac {\sqrt {v_{2}} \left(2v_{1}v_{2}(\Lambda
  v_{1}-2)+{e}^{2}v_{2}+12\,{ {v_{1}}}^{2} \right) {\pi
 }^{2}}{{8G_{5}}\,{v_{1}}}}. \label{Entropy_Function5D}
\end{eqnarray}
Since the system of equations of motion (\ref{R4_4}) near the
horizon are:
%
  \begin{eqnarray}
      0&=&{\frac {\pi \left( 3{v_{1}}+v_{2} \right) \left( 3{v_{1}}-{v_{2}} \right)  {a_{2}}}{\sqrt
       {v_{2}}{{v_{1}}}^{2}}}+{\frac {3\pi\left( 2\,{v_{1}}-{v_{2}} \right)
         \left( 2\, {v_{1}}+{v_{2}} \right){b_{2}}}{40\sqrt{v_{2}}{{v_{1}}}^{2}}}-{\frac {\pi \,\sqrt
          {{v_{2}}}\left(-12\,{{v_{1}}}^{2}-2\,\Lambda\,{{v_{1}}}^{2}{v_{2}}+{e}^{2}{v_{2}}\right)}
            {8{v_{1}}^2}},\nonumber \\
              0&=&{\frac{3\pi \left( {v_{1}}+{v_{2}} \right)  \left(3{v_{1}}
              -{v_{2}} \right) {a_{2}}}{\sqrt {{v_{2}}}{{v_{1}}}^{2}}}+{\frac
            {3\pi\left( 2\,{v_{1}}+{v_{2}} \right) \left( 2\, {v_{1}}-3\,{v_{2}}
           \right) {b_{2}}}{40\sqrt {v_{2}}{v_{1}}^{2}}}-{\frac {3\pi \,\sqrt
          {{v_{2}}} \left(2{v}^{2}_{1}(2+\Lambda v_{2})+v_{2}(e^2-4v_{1})
       \right) }{{{8v_{1}}}^{2}}},\nonumber \\
      0&=&q-\,{\frac {\pi \,{{v_{2}}}^{\frac32}e}{{v_{1}}\,{8G_{5}}}},
  \end{eqnarray}
this system can be solved explicitly, and then the functions
$e,v_{1}$ and $q$ can be written in terms of $v_{2}$ as:
%
 \begin{eqnarray}
  {v_{1}}&=&{\frac{{v_{2}}\left(5\,{v_{2}}+20{a_{2}}-{b_{2}}
      \right) }{60{a_{2}}+2\,{b_{2}}+5v_{2}(4+v_{2}\Lambda)}}, \label{v1_2ndOrder_5D} \\
      q&=&{\frac {\pi\,\sqrt{v_{2}}\,\widetilde{f}}{8\left( 5{v_{2}}+20{a_{2}}-{b_{2}} \right){G_{5}}}},\,\,\,
  e\, = \,\frac {\,\widetilde{f}}{20{v_{2}}+60{a_{2}}+2{b_{2}}+5{v}^{2}_{2}\Lambda}.
\end{eqnarray}
and,
$$
      \widetilde{f} = \sqrt {5}\,{\left( {{v_{2}}}^{2}\Lambda+20{a_{2}}+6{v_{2}}
    \right)}^{\frac{1}{2}}{\left(10{v}^{2}_{2}-60a_{2}b_{2}-40a_{2}{v}^{2}_{2}\Lambda-2{b}^{2}_{2}-3b_{2}{v}^{2}_{2}\Lambda-10b_{2}v_{2}
  \right)}^{\frac{1}{2}}. \nonumber
$$
Therefore, the entropy of a static and extremal BH in $d=5$, with
higher derivative terms of second degree taken into account, has the
form:
\begin{equation}
  {S}_{BH}={\frac {{\pi }^{2}{{v_{2}}}^{\frac{5}{2}} \left( 40{a_{2}}+3{b_{2} } \right)
   \Lambda}{ 4\left( -5{v_{2}}-20\,{a_{2}}+{b_{2}}\right) {G_{5}}}}+{\frac{5{\pi}^{2}
   {{v^{\frac{1}{2}}_{2}}}\left(-{{v_{2}}}^{2}+10{a_{2}}{b_{2}}+2{b_{2}}{v_{2}}\right)}{2\left(-5
   {v_{2}}-20{a_{2}}+{b_{2}} \right)
   {G_{5}}}}.\label{SBH_Second_Order_d5}
\end{equation}
Likewise, the extremal Gauss-Bonnet solution shown in
\cite{Morales:2006gm} can be obtained with the substitutions:
${b_{2}}=-8\,{\it \alpha}$, and $a_{2}=\frac{3}{5}{\it \alpha}$, in
the solution (\ref{SBH_Second_Order_d5}), leading to:
\begin{equation}
   S_{GB}=(v_{2}+12\alpha)\frac{\pi^2{v_{2}}^{\frac{1}{2}}}{2G_{5}}.\label{GB_5D}
\end{equation}
This does not happen for a more general GB contribution (see next
section). In general, establishing a relation between the GB's
solutions and the solutions of second degree requires a
constraint.\\ If $a_{2}=b_{2}=0$ in (\ref{SBH_Second_Order_d5}), the
well-known extremal RN solution is also obtained:
\begin{equation}
{S}_{BH}|_{a_{2}=b_{2}=0}=\frac{\pi^2{v}^{\frac{3}{2}}_{2}}{2G_{5}}\equiv
S^{d=5}_{RN}.
\end{equation}
As with the solution (\ref{SBH_Second_Order_d4}), the
cosmological constant does not change the entropy of a BH by itself: it
would need the higher derivative terms. A proof of this follows from
taking $\Lambda=0$ in (\ref{SBH_Second_Order_d5}), so that the contributions
of $a_{2}$ and $b_{2}$ remain. These contributions also produce a marked deviation
from the area law. Note that, contrary to the $d=4$ case, not only the
cosmological constant modifies the geometry of the BH throat (see
(\ref{v1_2ndOrder_5D})), but the coupling constants associated to the
invariants of second degree increase its effect on the throat
topology.
\subsection{Approximated Solutions for the complete set of invariants
in $d=5$. }
As in section \ref{Section-4}, we can construct approximated
solutions in five dimensions. Thus, considering the extremal BH
entropy in which the complete set of Riemann invariants (i.e. in
(\ref{AccionGeneralizada5D}) has been taken into account, the
coupling constant $a_{j},b_{j},c_{j},e_{j}\neq 0$) at different
levels of approximation take the form:
\begin{equation}
S_{BH}=
{S}^{d=5}_{RN}+(S^{(2)}_{BH}+S^{(3)}_{BH}w+S^{(4)}_{BH}w^2){\it
w}+{\it O(w^4)}.
\end{equation}
 where:
\begin{eqnarray}
S^{(2)}_{BH}&=& -\left( 2\,{{\left(1+\Lambda\,{v_{2}} \right)
{a_{2}}}}+{\frac{3}{20}}\,{{\left( 6+\Lambda\,{v_{2}} \right)
{b_{2}}}} \right)\frac{\pi^2{v^{\frac{1}{2}}_{2}}}{G_{5}}\,\,,\label{Approach_2nd_5D} \\
{S}^{(3)}_{BH}&=& \Biggl\{6\,{ { \left( 1+\Lambda\,{v_{2}} \right)
^{2}{a_{3}} }}+{\frac {3}{20}}\,{{\left( 6+\Lambda\,{v_{2}} \right)
\left( 3\,\Lambda\,{v_{2}}+8 \right) {b_{3}}}}-{\frac
{9}{400}}\,{{\left( 6+\Lambda\,{v_{2}} \right) ^{2}{c_{3}}}}+ \label{Approach_3nd_5D}\\
\nonumber &+& 8(1+\Lambda
v_{2})a_{2}^2+(-9+v_{2}\Lambda)\frac{a_{2}b_{2}}{5}-\frac{3}{100}(v_{2}\Lambda+6)b_{2}^2
\Biggr\}\frac{\pi^2}{{v^{\frac{1}{2}}_{2}}G_{5}},\\
{S}^{(4)}_{BH}&=& -\Biggl\{ {{{16}\left( 1+\Lambda\,{v_{2}}
\right)^{3}{a_{4}}}}+{{\frac9{100}\left( 6+\Lambda\,{v_{2}}
\right)^{3}{b_{4}}}}+{{\frac35\left( 6+\Lambda\,{v_{2}} \right)
\left( 2\,\Lambda\,{v_{2}}+7 \right)\left( 1+\Lambda\,{v_{2}}
\right){c_{4}}}}- \nonumber \\
&-&{{\frac3{200}\left( 4\,\Lambda\,{v_{2}}+9 \right)\left( 6+
\Lambda {v_{2}} \right)
^{2}{e_{4}}}}-32(1+v_{2}\Lambda)a_{2}^3+\frac{4}{5}(11+v_{2}\Lambda)b_{2}a_{2}^2+\label{Approach_4nd_5D}\\
\nonumber
&+&(9+4v_{2}\Lambda)\frac{a_{2}b_{2}^2}{25}-8(10+v_{2}\Lambda)(1+v_{2}\Lambda)^2a_{2}a_{3}
-(270+248v_{2}\Lambda+3v_{2}^3\Lambda^3+56v_{2}^2\Lambda^2)\frac{a_{2}b_{3}}{5}+\\
\nonumber
&+&\frac{3}{100}(v_{2}\Lambda+6)(v_{2}^2\Lambda^2+16v_{2}\Lambda+30)a_{2}c_{3}-\frac{3}{500}(v_{2}\Lambda+6)b_{2}^3-
\frac{3}{5}(1+v_{2}\Lambda)(v_{2}\Lambda+5)\times \\ \nonumber
&\times&(v_{2}\Lambda-4)a_{3}b_{2}-\frac{3}{200}(v_{2}\Lambda+6)(3v_{2}^2\Lambda^2+8v_{2}\Lambda-20)b_{2}b_{3}
+\frac{9 \Lambda v_{2}}{4000}(v_{2}\Lambda+6)^2 b_{2}c_{3}
\Biggr\}{\frac{\pi^2\,}{G_{5}{v_{2}}^{\frac32}}}\,.
\end{eqnarray}
%
 It is straightforward to notice that if equation
(\ref{SBH_Second_Order_d5}) is expanded in series up to second order
of $v^{-1}_{2}$, then equation (\ref{Approach_2nd_5D}) is
reproduced. If we take $a_{2}=\frac{3}{5}\alpha$ and
$b_{2}=-8\alpha$ in (\ref{Approach_2nd_5D}) the GB solution
(\ref{GB_5D}) is obtained. In this case, the non-linear
contributions of the coupling constants appear in $S^{(3)}_{BH}$ and
$S^{(4)}_{BH}$ as combinations of coupling constants of smaller
order like $a^{2}_{2}, a_{2}b_{2},b^{2}_{2},$ and
$a^{3}_{2},$...etc. Note that the absent terms
$a^{2}_{2},a_{2}b_{2}$ and $b^{2}_{2}$ in \ref{Approach_3nd_4D} in
contrast with \ref{Approach_3nd_5D} is simply a result of the
calculation. The same happen in \ref{Approach_4nd_4D} and
\ref{Approach_4nd_5D} with the terms,
$a^{3}_{2},b_{2}a^{2}_{2},a_{2}b^{2}_{2},b^{3}_{2}$.
\section{Riemann invariants and generic Gauss-Bonnet gravity.}
 In this section we examine the case of Einstein--Maxwell theory with a
slightly more general Gauss-Bonnet gravity ({\it i.e.} with three
different coupling constants). The action of this theory can be
written in $d$-dimensions as:
\begin{equation}
{\mathcal{S}}=\frac{1}{16\pi
      G_{d}}\int{dx^{d}\sqrt{-g}\left(R+\Lambda-\frac{F^{2}}{4}+\chi_{1}
      R_{\alpha \beta \gamma \delta}\,R^{\alpha \beta \gamma \delta}-4\,\chi_{2}R_{\mu \nu}\,R^{\mu \nu}+\chi_{3}R^{2}\right)}\,.
      \label{GB_with_3_coupling}
\end{equation}
so that, when the entropy function mechanism is applied, the entropy
of the extremal BH for $d=4$ takes the form: \\
%
\begin{eqnarray}
S^{(d=4)}_{GB}&=& -{\frac {2\pi \, \left(
\,{{v_{2}}}^{2}{\Lambda}^{2}+2\,{v_{2}}\, \Lambda+2 \right)
{\chi_{1}}}{{G_{4}}\, \left( {v_{2}}\,\Lambda+1
 \right) }}+{\frac {4\pi \, \left(2+\,{{v_{2}}}^{2}{\Lambda}^{2}+2
\,{v_{2}}\,\Lambda \right) {\chi_{2}}}{{G_{4}}\, \left( {v_{2}}\,
\Lambda+1 \right) }},  \label{Entropia_GB_3CC_4D}\\
&-& \,{\frac {2\pi \,{{v_{2}}}^{2}{\Lambda}^{2}{\chi_{3}}}{{G_{4}}\,
\left( {v_{2}}\,\Lambda+1 \right) }}+\frac{\pi\,v_{2}}{G_{4}}\,.
\nonumber
\end{eqnarray}
The solutions for the functions $v_{1},\,e,$ and $q$ are:
\begin{eqnarray}
  v_{1}&=& \frac{v_{2}}{v_{2}\,\Lambda+1},\qquad e=\frac{{f_{0}}}{v_{2}\,\Lambda+1},
   \qquad q=-\frac{{f_{0}}}{4G_{4}},  \\
    f_{0}&=& {( {{{-8{ \chi_{1}}\,{{v_{2}}}^{2}{\Lambda}^{2}-16\,
     {\chi_{1}}\,{v_{2}}\,\Lambda+16\,{\chi_{2}}{{v_{2}}}^{2}{\Lambda}^
     {2}+32\,{\chi_{2}}\,{v_{2}}\Lambda-8{\chi_{3}}{{v_{2}}}^{2}{
   \Lambda}^{2}-16\,{\chi_{3}}{v_{2}}\Lambda+2{{v_{2}}}^{2}
  \Lambda+4{v_{2}}-{{p}}^{2}}}} )}^\frac{1}{2}\,.\nonumber
\end{eqnarray}
so that, if $\chi_{1}=\chi_{2}=\chi_{3}=\chi$, in
(\ref{Entropia_GB_3CC_4D}), the solutions (\ref{GB_4D}) is obtained
(with $\chi=\alpha$). However, if we want to find the general GB
solutions (\ref{Entropia_GB_3CC_4D}) from the analytical solution
(\ref{SBH_Second_Order_d4})(or vice versa), then the following
constraints should be applied on both results:
\begin{equation}
 \chi_{1}-2\chi_{2}+\chi_{3}=0, \qquad
   \chi_{1}=2\chi_{2}+\frac{3b_{2}}{16}+a_{2}, \qquad
 \chi_{3}=a_{2}-\frac{b_{2}}{16}\,.
\end{equation}
Likewise, for $d=5$ the resultant entropy is given by,
\begin{eqnarray}
 S^{(d=5)}_{GB}=
   \frac{\pi^2}{(v_{2}+4\chi_{3})G_{5}}(-8\sqrt{v_{2}}\chi^{2}_{1} &+&
    \chi_{1}(\,48\chi_{2}\sqrt{v_{2}}-
    2\sqrt{v_{2}}(4v_{2}+16\chi_{3}+\Lambda\,v^{2}_{2}))-64\sqrt{v_{2}}\chi^{2}_{2}+\, \nonumber \\
   +\,4\,\sqrt{v_{2}}(\,20\chi_{3} &+& \Lambda v^{2}_{2}+4v_{2})\chi_{2}-\frac{v^{\frac{5}{2}}_{2}}{2}(4\chi_{3}\Lambda-1)
 )\,.   \label{Entropia_GB_3CC_5D}
\end{eqnarray} 
for the funtions $v_{1},\,e$ and $q$:
\begin{eqnarray}
v_{1}&=&\frac{v_{2}(v_{2}+4\chi_{3})}{4v_{2}+\Lambda
v^{2}_{2}-16\chi_{2}+12\chi_{3}+4\chi_{1}}\,,
\\
q\, &=& \,q(v_{2},\Lambda,\chi_{1},\chi_{2},\chi_{3}),\,\qquad
e\,=\,e(v_{2},\Lambda,\chi_{1},\chi_{2},\chi_{3})\,.
\end{eqnarray}
%
Also, the relation between the entropies (\ref{Entropia_GB_3CC_5D})
and (\ref{SBH_Second_Order_d5}) are given for the case when the
following constraints hold:
\begin{equation}
\chi_{1}-2\chi_{2}+\chi_{3}=0, \qquad
a_{2}=\frac{3}{5}(2\chi_{2}-\chi_{1}), \qquad
b_{2}=8(\chi_{1}-2\chi_{2})\,.
\end{equation}
 If we take $\,\chi_{1}=\chi_{2}=\chi_{3}=\chi$, in (\ref{Entropia_GB_3CC_5D}), equation (\ref{GB_5D}) is obtained.
 These last results are interesting because they correspond to
cases where the Riemann invariants cannot reproduce the results of
the generic GB theory, and this suggests that this theory will
always lack invariants to add. However, this situation can be
understood if we notice that it is a well-known fact that generic GB
gravity in AdS gravity is, in general, inconsistent by two essential
points: the variational principle \cite{Aros:1999id}, and
regularization problems, both subjects are discussed extensively in
\cite{Olea:2005gb},\cite{Kofinas:2006hr},\cite{Kofinas:2008ub}.
\section{Conclusion.}
We have calculated the entropy for extremal BHs in the cases  $d=4$
and $d=5$, taking into account higher derivative terms built from
the complete set of Riemann invariants, a task which, as far as we
are aware, has not been accomplished before. Though we remark that
the exceptional Gauss-Bonnet's cases were already done in
\cite{Morales:2006gm}. We have found the invariants of second degree
and generalized these results so that the GB case can be obtained as
a particular case. We have also obtained the leading terms of
approximation for higher order invariants that could be interpreted
as contributions of higher derivative terms in some theories of
gravity. The RN solutions are also contained in all the examined
cases. Therefore, the use of our set of solutions provides a
concrete example showing that the entropy function formalism works
well, and its applications can be a less complicated process in
comparison to the use of Wald's equation
\cite{Matyjasek:2008yq}-\cite{Matyjasek:2006nu}.
\section*{Acknowledgements}
{This work was partially supported by {\it ICTP Federation
Arrangement Program} of Trieste, Italy and  {\it INFN Universita di
Roma Tor Vergata} in Roma, Italy. The author is indebted to Jose. F.
Morales for his useful discussions and original ideas.
Acknowledgement is also due to Roberto A. Sussman, A. Cabo Montes de
Oca, A. P\'{e}rez Mart\'{\i}nez and H. P\'{e}rez Rojas for their
support and interest in the elaboration of this work.}
%
%
%
\appendix
\begin{table}[!ht]
\caption{\label{Definicion_de_Invariantes} Table of definitions for
the complete set of Riemann invariants and the results for the
$AdS_{2}\times S^{d-2}$ geometry near the BH horizon in $d=4$ and
$d=5$ (the rest of the invariants are zero). All the definitions are
based on the trace--free Ricci tensor $S_{a\,b}$, the Weyl tensor
$C_{abcd}$ and the Riemann tensor $R_{abcd}$. Although, these
invariants can also be defined on a spinor base \cite{JCarminati}.
Here ``-'' means that there is no invariant definition available in
five dimensions for the set of complex invariants. We use small
letters for the invariants of Carminati \cite{JCarminati}, but in
the Lagrangean we use upper case letters. The $\mathfrak{Re}$ symbol
means real part of the complex invariant. In order to remark the
tensor degree (instead of Carminati subindex) we have written the
tensor degree number as a subindex ({\it i.e.}
$\mathfrak{Re}(W_{2})$ means the real part of the second degree
complex Weyl invariant $w_{1}$).}
\begin{tabular}{@{}|l|l|l|l|}
\hline
  $ $ & $ $ & $ $ & $$ \\
    $\,Invariants $ & $ \,Definitions $ & $\,\, Solutions $ & $Solutions  $ \\
  $ $ & $ $ & $\,d=4$ & $\,d=5 $ \\
\hline
    $ $ & $ $ & $ $ & $$ \\
    $ R $ & $g^{ad}g^{bc}R_{abcd} $ & $\,2\,\gamma_{1} $ & $2\gamma_{3}$ \\
    \cline{1-4}
     $ $ & $ $ & $ $ & $$ \\
        $r_{1}=R_{2} $ & $\frac{1}{4}S_{a}^{\,\,\,b}S_{b}^{\,\,\,a} $ & $\frac{1}{4}\gamma^{2}_{2} $ & $\frac{3}{10}\gamma^{2}_{4}$ \\
    \cline{1-4}
     $ $ & $ $ & $ $ & $$ \\
        $r_{2}=R_{3} $ & $-\frac{1}{8}S_{a}^{\,\,b}S_{b}^{\,\,c}S_{c}^{\,\,a}$ & $0 $ & $\frac{3}{10^2}\gamma^{3}_{4}$\\
    \cline{1-4}
        $ $ & $ $ & $$ & $$  \\
        $r_{3}=R_{4} $ & $\frac{1}{16}S_{a}^{\,\,b}S_{b}^{\,\,c}S_{c}^{\,\,d}S_{d}^{a} $ & $\frac{1}{64}\gamma^{4}_{2} $ & $\frac{21}{10^3}
        \gamma^{4}_{4} $ \\
\hline
 $ $ & $ $ & $ $ & $$ \\
       ${\mathfrak{Re}}(w_{1})=\mathfrak{Re}(W_{2}) $  & $\frac{1}{8}C_{abcd}C^{\,abcd} $ & $\frac{1}{6}\gamma^{2}_{1} $ & $\hspace{0.5cm}-$ \\
     \cline{1-4}
      $ $ & $ $ & $ $ & $$ \\
       ${\mathfrak{Re}}(w_{2})=\mathfrak{Re}(W_{3})$ & $-\frac{1}{16}C_{ab}^{\,\,cd}C_{cd}^{\,\,ef}C_{ef}^{\,\,ab} $ & $\frac{-1}{36}\gamma^{3}_{1}$ & $\hspace{0.5cm}-$ \\
     \cline{1-4}
      $ $ & $ $ & $ $ & $$ \\
       $\mathfrak{Re}(m_{1})=\mathfrak{Re}(M_{3})$ & $\frac{1}{8}S^{ab}S^{cd}C_{acdb} $ & $\frac{-1}{12}\gamma_{1}\gamma^{2}_{2} $ & $\hspace{0.5cm}-$ \\
   \cline{1-4}
    $ $ & $ $ & $ $ & $$ \\
       $\mathfrak{Re}(m_{2})=\mathfrak{Re}(M_{4})$ & $\frac{1}{16}S^{cd}S_{ef}(C_{acdb}C^{aefb}-C^{*}_{acdb}C^{*aefb}) $ & $\frac{1}{36}\gamma^{2}_{1}
       \gamma^{2}_{2} $ & $\hspace{0.5cm}-$ \\
     \cline{1-4}
      $ $ & $ $ & $ $ & $$ \\
       $m_{3}={M^{+}_{4}}$ & $\frac{1}{16}S^{cd}S_{ef}(C_{acdb}C^{aefb}+C^{*}_{acdb}C^{*aefb}) $ & $\frac{1}{36}\gamma^{2}_{1}
       \gamma^{2}_{2}$ & $\hspace{0.5cm}-$  \\
     \cline{1-4}
 \cline{1-4}
  $ $ & $ $ & $ $ & $$ \\
       $\mathfrak{Re}(m_{5})=\mathfrak{Re}(M_{5})$ & $\frac{1}{32}S^{cd}S^{ef}C^{aghb}(C_{acdb}C_{gefh}+C^{*}_{acdb}C^{*}_{gefh}) $ & $\frac{-1}{108}\gamma^{3}_{1}
       \gamma^{2}_{2}$ & $\hspace{0.5cm} -$  \\
     \cline{1-4}
      $\hspace{1.7cm}\,where:$ & $\gamma_{1}=(v_{1}-v_{2})/v_{1}v_{2},\,\,\gamma_{2}=(v_{1}+v_{2})/v_{1}v_{2},$ & $ $ & $$ \\
      $ $ & $\,\gamma_{3}=(3v_{1}-v_{2})/v_{1}v_{2},\,\gamma_{4}=(2v_{1}+v_{2})/v_{1}v_{2} $ & $ $ & $$ \\
\hline
\end{tabular}
\end{table}
\begin{table}[!ht]
\caption{\label{Tabla_de_Invariantes} The complete set of not null
invariants is provided, organized by $degrees$ for $d=4$ and $d=5$
dimensions on $AdS_{2}\times S^{d-2}$ geometry.}
\begin{tabular}{@{}|l|l|l|l|}
    \hline
         $ $ & $ $ & $ $ & $$ \\
         $\,d $ & $ \,Degree $ & $\,\, Invariants $ & $Equivalence $ \\
         $ $ & $ $ & $ $ & $relations $ \\
    \hline
         $ d=4 $ & $ 1^{\underline{st}} $ & $ R=2\gamma_{1} $ & $\,\,\gamma_{1} $ \\
    \cline{2-4}
         $ $ & $2^{\underline{nd}} $ & $R^{2}=4\gamma^{2}_{1},\qquad R_{2}=\frac{1}{4}{\gamma}^{2}_{2}, \qquad \mathfrak{Re}(W_{2})=
         \frac{1}{6}{\gamma}^{2}_{1}, $
         & $\gamma^{2}_{1},\,\gamma^{2}_{2}$ \\
    \cline{2-4}
        $ $ & $3^{\underline{rd}} $ & $R^{3}=8\gamma^{3}_{1},\qquad R\times R_{2}=\frac{1}{2}\gamma_{1}\gamma^{2}_{2},\qquad R\times \mathfrak{Re}(W_{2})=\frac{1}
        {3}\gamma^{3}_{1},\,$ & $\,\,\gamma^{3}_{1},\,
        \gamma_{1}\gamma^{2}_{2}$\\
        $ $ & $ $ & $\mathfrak{Re}(W_{3})=\frac{-1}{36}{\gamma}^{3}_{1},\qquad \mathfrak{Re}(M_{3})=\frac{-1}{12}\gamma_{1}{\gamma}^{2}_{2} $ & $ $ \\
    \cline{2-4}
        $ $ & $ $ & $ $ & $ $ \\
        $ $ & $4^{\underline{th}} $ & $R^{4}=16\gamma^{4}_{1}, \qquad \mathfrak{Re}^2({W_{2}})=\frac{\gamma^{4}_{1}}{36}, \qquad
         R^{2}\times R_{2}=\gamma^{2}_{1} \gamma^{2}_{2},$ & $ $ \\
         $ $ & $ $ & $ \qquad R_{2}^{2}=\frac{1}{16}\gamma^{4}_{2}, \qquad R^{2}\times \mathfrak{Re}(W_{2})=\frac{2}{3}\gamma^{4}_{1},  $ & $ $ \\
        $ $ & $ $ & $ R_{2}\times \mathfrak{Re}(W_{2})=\frac{1}{24}\gamma^{2}_{1}\gamma^{2}_{2},\, \qquad \mathfrak{Re}(M_{4})=M^{+}_{4}=\frac{1}{36}
        {\gamma}^{2}_{1}{\gamma}^{2}_{2},  $ & $\,\gamma^{4}_{1},\,\gamma^{4}_{2},\,\gamma^{2}_{1}\gamma^{2}_{2}$   \\
        $ $ & $ $ & $\qquad R\times  \mathfrak{Re}(W_{3})=\frac{-\gamma^{4}_{1}}{18},
        \qquad R\times \mathfrak{Re}(M_{3})=\frac{-\gamma^{2}_{1}\gamma^{2}_{2}}{6}, $ & $ $ \\
        $ $ & $ $ & $ R_{4}=\frac{1}{64}{\gamma}^{4}_{2},$ & $ $\\
    \cline{2-4}
        $ $ & $ $ & $ $ & $ $ \\
        $ $ & $5^{\underline{th}} $ & $R^5={{32}}\gamma^{5}_{1}, \qquad R^3\times R_{2}=2\gamma^{3}_{1}\gamma^{2}_{2}, \qquad R^3\times
        \mathfrak{Re}(W_{2})=\frac{4}{3}\gamma^{5}_{1}, $ & $ $ \\
        $ $ & $ $ & $\qquad R^2\times \mathfrak{Re}(W_{3})=-\frac{\gamma^{5}_{1}}{9}, \qquad R^{2}_{2}\times{R}=\frac{\gamma_{1}\gamma^{4}_{2}}{8},$ & $ $ \\
        $ $ & $ $ &  $ R^2\times \mathfrak{Re}(M_{3})=-\frac{\gamma^{3}_{1}\gamma^{2}_{2}}{3},\qquad R\times \mathfrak{Re}^{2}(W_{2})=
        \frac{\gamma^{5}_{1}}{18},$ & $\gamma^{5}_{1},\gamma_{1}\gamma^{4}_{2},\gamma^{3}_{1}\gamma^{2}_{2}$ \\
        $ $ & $ $ & $\qquad \,R\times R_{2}\times \mathfrak{Re}(W_{2})=
        \frac{\gamma^{3}_{1}\gamma^{2}_{2}}{12}, $ & $ $ \\
        $ $ & $ $ &  $R_{2}\times \mathfrak{Re}(W_{3})=\frac{-\gamma^{3}_{1}\gamma^{2}_{2}}{144}, \qquad R_{2}\times
        \mathfrak{Re}(M_{3})=\frac{-\gamma_{1}\gamma^{4}_{2}}{48}, \qquad $ & $ $ \\
        $ $ & $ $ & $ \mathfrak{Re}(W_{2})\times \mathfrak{Re}(M_{3})=\frac{-\gamma^{3}_{1}\gamma^{2}_{2}}{72},\qquad
         \mathfrak{Re}(W_{2})\times \mathfrak{Re}(W_{3})=\frac{-\gamma^{5}_{1}}{216}, $ & $ $ \\
        $ $ & $ $ & $ R\times R_{4}=\frac{\gamma_{1}\gamma^{4}_{2}}{32}, \qquad R\times \mathfrak{Re}(M_{4})=R\times M^{+}_{4}\equiv\frac{\gamma^{3}_{1}\gamma^{2}_{2}}{18}$ & $ $ \\
\hline
       $ $ & $ $ & $ $ & $ $ \\
       $ d=5 $  & $1^{\underline{st}} $ & $R=2\gamma_{3} $ & $ \, \gamma_{3}$ \\
     \cline{2-4}
       $ $ & $2^{\underline{nd}} $ & $R^{2}=4\gamma^{2}_{3}, \qquad R_{2}=\frac{3}{10}\gamma^{2}_{4}\qquad$ & $\gamma^{2}_{3},\,\gamma^{2}_{4}$ \\
     \cline{2-4}
       $ $ & $3^{\underline{rd}} $ & $R^{3}=8\gamma^{3}_{3},\qquad R\times\,R_{2}=\frac{3}{5}\gamma_{3}\gamma^{2}_{4},\qquad R_{3}=\frac{3}{100}\gamma^{3}_{4} $
       & $\gamma^{3}_{3},\,\gamma_{3}\gamma^{2}_{4},\,\gamma^{3}_{4}$ \\
     \cline{2-4}
       $ $ & $4^{\underline{th}} $ & $R^{4}=16\gamma^{4}_{3},\qquad R^{2}_{2}=\frac{9}{100}\gamma^{4}_{4},\qquad R^{2}\times\,
       R_{2}=\frac{6}{5}\gamma^{2}_{3}\gamma^{2}_{4}, \qquad
        $ & $\,\gamma^{4}_{3},\,\gamma^{4}_{4},\,\gamma^{2}_{3}\gamma^{2}_{4},$  \\
        $ $ & $ $ & $\qquad R\times R_{3}=\frac{3}{50}\gamma_{3}\gamma^{3}_{4}, \qquad R_{4}=\frac{21}{10^3}\gamma^{4}_{4}$ & $\gamma_{3}\gamma^{3}_{4} $ \\
     \cline{2-4}
       $ $ & $ $ & $where:\,\,\,\gamma_{1}=(v_{1}-v_{2})/v_{1}v_{2},\,\gamma_{2}=(v_{1}+v_{2})/v_{1}v_{2}, $ & $$ \\
       $ $ & $ $ & $\hspace{1.3cm}\gamma_{3}=(3v_{1}-v_{2})/v_{1}v_{2},\,\gamma_{4}=(2v_{1}+v_{2})/v_{1}v_{2} $ & $$ \\
\hline
\end{tabular}
\end{table}
\newpage
%

%

\begin{thebibliography}{999}
%
\bibitem{Sen:2007qy}
  A.~Sen,
  ``Black Hole Entropy Function, Attractors and Precision Counting of
  Microstates,''\,{\it Preprint:}\,[\arXivid{0708.1270},\,[hep-th]].
%
\bibitem{Examples_EF_1} A. Sen, JHEP 0509, 038 (2005) [arXiv:\hepth{0506177}]; A. Sen,
JHEP 0603, 008 (2006) [arXiv:\hepth{0508042}].
%
\bibitem{Examples_EF_2} A. Ghodsi, Phys. Rev. D 74, 124026 (2006)
[arXiv:\hepth{0604106}].
%
 \bibitem{Examples_EF_3} P. Prester, JHEP 0602, 039 (2006)
[arXiv:\hepth{0511306}]. A. Sinha and N. V. Suryanarayana, Class.
Quant. Grav. 23, 3305 (2006) [arXiv:\hepth{0601183}]. B. Sahoo and
A. Sen, JHEP 0609, 029 (2006) [arXiv:\hepth{0603149}]. G. Exirifard,
JHEP 0610, 070 (2006) [arXiv:\hepth{0604021}]. B. Chandrasekhar,
arXiv:hep-th/0604028. R. G. Cai and D. W. Pang, Phys. Rev. D 74,
064031 (2006) [arXiv:\hepth{0606098}]. A. Sinha and N. V.
Suryanarayana, JHEP 0610, 034 (2006) [arXiv:\hepth{0606218}]. B.
Sahoo and A. Sen, JHEP 0701, 010 (2007) [arXiv:\hepth{0608182}]. G.
L. Cardoso, J. M. Oberreuter and J. Perz, arXiv:\hepth{0701176}. K.
Goldstein and R. P. Jena, arXiv:\hepth{0701221}. R. G. Cai and D. W.
Pang, arXiv:\hepth{0702040}. M. Alishahiha, arXiv:\hepth{0703099},
R. G. Cai, C. M. Chen, K. i. Maeda, N. Ohta and D. W. Pang,
Phys.\,Rev.\,D\,{\bf 77},\, 064030\,(2008)\,
[\arXivid{0712.4212},\,[hep-th]].
%
\bibitem{James1}
Pau Figueras,\,Hari K Kunduri,\,James Lucietti, and \,Mukund
Rangamani, Extremal vacuum black holes in higher dimensions,\,{\it
Phy.\,Rev.\,D} {\,\bf 78}: 044042,\,
(2008)\,[\arXivid{0803.2998},\,[hep-th]].
%
\bibitem{James2}
Hari K Kunduri,\,James Lucietti,\, and Harvey S.
Reall,\,Near-horizon symmetries of extremal black holes,\,{\it
Class.\,Quant.\,Grav},\,{\bf
24}:\,4169-4190,\,(2007)\,\,[\arXivid{0705.4214},\,[hep-th]].
%
\bibitem{Astefanesei1}
Dumitru Astefanesei and Hossein Yavartanoo, Stationary black holes
and attractor mechanism,\,{\it Nucl.\,Phys.\,B,\,} {\bf
794}:\,13-27, (2008),\,[\arXivid{0706.1847},\,[hep-th]].
%
\bibitem{Robinson} I. Robinson \,1959\,A Solution of the
Maxwell-Einstein Equations, {\it Bull.\,Acad.\,Polon.\,Sci.}\,
{\textbf{7}},\, 351-352.
%
\bibitem{Bertotti} B.\,Bertotti \,Uniform Electromagnetic Field in the Theory of
General Relativity \pr{116}{1959}{1331}.
%
\bibitem{T.Ortin} Tomas Ortin \,2004 \,Agujeros negros clasicos y
cuanticos en Teoria de Cuerdas
\,{\it{Preprint:}}\,\,[arXiv:\hepth{0405005}].
%
\bibitem{Gibbons}G.W.Gibbons, \, 1985 \,Aspects of Supergravity Theories, (three
lectures) in: Supersymmetry, Supergravity and Related Topics, eds.F.
del Aguila, J. de Azcarraga and L.Iba{\~n}ez, World Scientific,
Singapore, p.147.
%
\bibitem{Farhoudi}Mehrdad Farhoudi, \, 2005 On Higher Order Gravities, Their Analogy to GR,
and Dimensional Dependent Version of Duffs Trace Anomaly
Relation.\grg{38}{2006}{1261} \,{\it{Preprint:}}\,\,
[arXiv:physics/0509210v2].
%
\bibitem{B.Zwiebach} B.Zwiebach, \, 1985 Curvature Squared Terms and
String Theories, {\it Phys. Lett. B},\,{\bf 156},\,315.
%
\bibitem{Morales:2006gm}
  J.~F.~Morales and H.~Samtleben,
  ``Entropy function and attractors for AdS black holes,''
  \jhep{0610}{2006}{074},
  [arXiv:\hepth{0608044}].
%
\bibitem{Astefanesei2}
Dumitru Astefanesei,\,Nabamita Benerjee and Suvankar
Dutta,\,``(Un)attractor black holes in higher derivative AdS
gravity,'',[\arXivid{0806.1334},\,[hep-th]]
%
\bibitem{Astefanesei3}
  D.~Astefanesei, H.~Nastase, H.~Yavartanoo and S.~Yun,
  ``Moduli flow and non-supersymmetric AdS attractors,''
  \jhep{0804}{2008}{074}
  [\arXivid{0711.0036} [hep-th]].
%
\bibitem{Maldacena:1996ky}
  J.~M.~Maldacena,
  ``Black holes in string theory,''
  arXiv:\hepth{9607235}.
%
\bibitem{Duff:1999rk}
  M.~J.~Duff,
  ``Lectures on branes, black holes and anti-de Sitter space,''
  arXiv:\hepth{9912164}.
%
\bibitem{Peet:2000hn}
  A.~W.~Peet,
  ``TASI lectures on black holes in string theory,''
  [arXiv:\hepth{0008241}].
%
\bibitem{JCarminati} Carminati.\,J and McLenaghan.\,R.G
\,1991\,Algebraic invariants of the Riemann tensor in a
four-dimensional Lorentzian space {\it{J.Math.Phys}}\,
Vol.{\textbf{32}},\,\,No.{11},\,3135-3140.
%
\bibitem{Zakhary_McIntosh} Zakhary.\,E and McIntosh.C.B.G \,1997\,A complete set of Riemann
invariants {\it{Gen.\,Rel.\,Grav}}\,
Vol.\,{\textbf{29}},\,\,No.{5},\, 539-581.\grg{29}{1997}{539}.
%
\bibitem{Matyjasek:2008yq}
  J.~Matyjasek,
  ``Entropy of nonlinear black holes in quadratic gravity,''
  Acta Phys.\ Polon.\  B {\bf 39} (2008) 3
  [\arXivid{0802.4063} [gr-qc]].
%
\bibitem{Matyjasek:2006fq}
  J.~Matyjasek, M.~Telecka and D.~Tryniecki,
  ``Higher dimensional black holes with a generalized gravitational action,''
  Phys.\ Rev.\  D {\bf 73} (2006) 124016
  [arXiv:\hepth{0606254}].

\bibitem{Berej:2006cc}
  W.~Berej, J.~Matyjasek, D.~Tryniecki and M.~Woronowicz,
  ``Regular black holes in quadratic gravity,''
  Gen.\ Rel.\ Grav.\  {\bf 38} (2006) 885
  [arXiv:\hepth{0606185}].
%
\bibitem{Matyjasek:2006nu}
  J.~Matyjasek,
  ``Entropy of quantum-corrected black holes,''
  {Phys.\ Rev.\  D} {\bf 74} (2006) 104030
  [arXiv:\grqc{0610020}].
  %
\bibitem{Aros:1999id}
  R.~Aros, M.~Contreras, R.~Olea, R.~Troncoso and J.~Zanelli,
  ``Conserved charges for gravity with locally AdS asymptotics,''
  Phys.\ Rev.\ Lett.\  {\bf 84} (2000) 1647
  [arXiv:\grqc{9909015}].
  %
\bibitem{Olea:2005gb}
  R.~Olea,
``Mass, angular momentum and thermodynamics in four-dimensional
  Kerr-AdS black holes,''
  JHEP {\bf 0506} (2005) 023
  [arXiv:\hepth{0504233}].
%
\bibitem{Kofinas:2006hr}
  G.~Kofinas and R.~Olea,
  ``Vacuum energy in Einstein-Gauss-Bonnet AdS gravity,''
  Phys.\ Rev.\  D {\bf 74} (2006) 084035
  [arXiv:\hepth{0606253}].
%
\bibitem{Kofinas:2008ub}
  G.~Kofinas and R.~Olea,
  ``Universal Kounterterms in Lovelock AdS gravity,''
  Fortsch.\ Phys.\  {\bf 56} (2008) 957
  [\arXivid{0806.1197}\,[hep-th]].
%





%
\end{thebibliography}
\end{document}